\documentclass[preprint,pteplogo]{ptephy}
\usepackage[utf8]{inputenc}
\usepackage{latexsym,array,theorem,mathrsfs,bm,float}
\usepackage{psfrag}
\usepackage{amsfonts,amsmath,amssymb,latexsym,array,afterpage,
theorem,mathrsfs,bm,float,epsfig,color,graphicx,tabularx,here,multirow}

\newcommand{\nn}{\nonumber \\}
\newcommand{\bea}{\begin{eqnarray}}
\newcommand{\ena}{\end{eqnarray}}
\newcommand{\beann}{\begin{eqnarray*}}
\newcommand{\enann}{\end{eqnarray*}}

\newcommand{\ma}[1]{\mbox{$\mathcal{#1}$}}

\newcommand{\calhR}[1]{\raisebox{2ex}{\tiny ({\em h})}\hspace{-0.8em}{\ma R}}

\newcommand{\mpl}{M_{\mathrm{PL}}}

\begin{document}

\title{A Possible Solution to the Helium Anomaly of EMPRESS VIII by Cuscuton Gravity Theory}

\author{\name{Kazunori \surname{{\sc Kohri}}}{1,2,3\ast},  
and 
\name{Kei-ichi \surname{\sc Maeda}} {4,5\ast\ast}
%\thanks{These authors contributed equally to this work}
}
%%%%%%%%%%% The \name command should be used as \name{Insert author name here}{Insert affiliation number here}
%%%%% Please use \thanks for contributed author details

%%%%%%%%%%% The \affil command should be used as \affil{Insert affiliation number here}{Insert author address here}
\address{
\affil{1}{Theory Center, IPNS, KEK, 1-1 Oho, Tsukuba 305-0801, Japan}
\affil{2}{Sokendai, 1-1 Oho, Tsukuba 305-0801, Japan}
\affil{3}{Kavli IPMU (WPI), UTIAS, The University of Tokyo, Kashiwa, Chiba 277-8583, Japan}
\affil{4}{Department of Pure and Applied Physics, Graduate School of Advanced Science and Engineering, Waseda University, 
Okubo 3-4-1, Shinjuku, Tokyo 169-8555, Japan}
\affil{5}{Center for Gravitational Physics and Quantum Information, Yukawa
Institute for Theoretical Physics, Kyoto University, Kyoto 606-8555, Japan}
\email{kazunori.kohri-at-gmail.com}\\
$^\ast$\email{maeda-at-waseda.jp}
}

\begin{abstract}
We discuss cosmology based on the cuscuton
gravity theory to resolve the anomaly of the observational $^4$He
abundance reported by the EMPRESS collaboration.
We find that 
the gravitational constant $G_{\rm cos}$ in Friedmann equation should be smaller than the Newton's constant ${G_{\rm N}}$ such that 
${\Delta G_{\rm N}}/{G_{\rm N}} \equiv (G_{\rm cos}-G_{\rm N})/{G_{\rm N}} = -0.085_{-0.028}^{+0.026}  \quad(68
\% \text { C.L. })$  in terms of big-bang
nucleosynthesis, which excludes ${\Delta G_{\rm N}}=0$ at more than
95~$\% \text { C.L. }$ To fit the data, we obtain a negative mass squared 
of  a non-dynamical scalar field with the Planck-mass scale as
$\sim - {\mathcal{O}}(1) {M_{\rm PL}^2} ({\mu}/{0.5 M_{\rm PL}})^{4}$ 
with the cuscuton mass parameter $\mu$. 
This fact could
suggest the need for modified gravity theories such as the cuscuton gravity
theory with a quadratic potential, which can be regarded as the low-energy Ho\v{r}ava-Lifshitz gravity,
and might give a hint of  quantum gravity.
\end{abstract}

%\subjectindex{xxxx, xxx}

\maketitle

%\end{titlepage}

%%%%%%%%%%%%%%%%%%%%%%%%%%%%%%%%%%%%%%%%%%%%%%%%%%%%%%%%%%%%%%%%%%%%%%
\section{Introduction}
%%%%%%%%%%%%%%%%%%%%%%%%%%%%%%%%%%%%%%%%%%%%%%%%%%%%%%%%%%%%%%%%%%%%%%
\label{sec:intro}

Quite recently, the EMPRESS collaboration (EMPRESS VIII) newly
observed 10 Extremely Metal-Poor Galaxies (EMPGs) with metallicity
($ < 0.1Z_{\odot}$) by using Subaru telescope, and obtaining data of
the $^4$He to hydrogen ratio ($^4$He/H) by measuring the He{\sc
  i}$\lambda$10830 near-infrared emission~\cite{Matsumoto:2022tlr}. By
analyzing the data of 64 galaxies in total with 13 EMPGs (including
the new 10 EMPGs), they estimated the primordial mass fraction of
$^4$He to be ${\rm Y}_p = 0.2379^{+0.0031}_{-0.0030},$ by
extrapolating the data into the value at zero metallicity (the oxygen
to hydrogen ratio O/H $\to$ 0).

Comparing the data with theoretical predictions in the standard
big-bang nucleosynthesis (BBN),  they obtained the effective number of
the neutrino species to be $N_{\nu, {\rm eff}} = 2.41^{+0.19}_{-0.21}$
at 68 $\%$ C.L. It is remarkable that this means that the standard
value of $N_{\nu, {\rm eff}}$ predicted in the big-bang cosmology (=
$N_{\nu, {\rm eff, std}} \simeq 3.044 -
3.046$~\cite{Mangano:2005cc,deSalas:2016ztq,EscuderoAbenza:2020cmq,Akita:2020szl,Froustey:2020mcq,Bennett:2020zkv}),
is observationally excluded at more than 2 $\sigma$.

Seriously-taking this discrepancy of $N_{\nu, {\rm eff}}$ in the
standard big-bang cosmology, we must consider modified theories beyond
the standard model. The EMPRESS collaboration extended their framework
to a new theory beyond the standard model by adding one more free
parameter, so-called ``degeneracy parameter'', $\xi_{\nu_e}$ which
means a non-zero lepton number in the electron neutrino sector. It has
been known for a long time that a positive $\xi_{\nu_e}$ can reduce
$Y_p$ without changing the number of neutrino species
much~\cite{Kohri:1996ke} by which the Helium anomaly can be solved
this time. The best-fit value is $\xi_{\nu_e} \sim 0.05$ with
excluding $\xi_{\nu_e}=0$ at more than 1
$\sigma$~\cite{Matsumoto:2022tlr} (See also~\cite{Burns:2022hkq,Escudero:2022okz} for a
similar analysis with detailed discussions about dependences on
nuclear-reaction rates in BBN). Theoretically, such a large lepton
number can be produced even after the cosmic temperature is smaller
than the weak scale ${\mathcal{O}}(10^2)$~GeV in models with $Q$-balls
($L$-balls)~\cite{Kawasaki:2002hq,Kawasaki:2022hvx}, late-time resonant
leptogenesis~\cite{Borah:2022uos}, oscillating
sterile neutrinos~\cite{Shi:1999kg}, etc. In future, we can measure
$\xi_{\nu_e}$ more precisely by planned observations of 21cm + CMB
down to errors of
$\Delta \xi_{\nu_e} \sim 5 \times 10^{-3}$~\cite{Kohri:2014hea}.

There is another way to solve this anomaly, which is a modification of the Einstein gravity. 
In terms of a modified gravity theory, recently an interesting model
to realize dark energy has been proposed, called a cuscuton gravity
theory~\cite{Afshordi:2006ad,Afshordi:2007yx} or its extended
version~\cite{Iyonaga:2018vnu,Iyonaga:2020bmm,Maeda:2022ozc}. In the context of the
beyond Horndeski theories, the original cuscuton gravity theory was
extended to be a generalized one~\cite{Gleyzes:2014dya}, in which the
second-order time derivatives of a scalar field in the equation of
motion disappears. Thus the scalar field appearing in the theories is
just a non-dynamical shadowy mode. 
There is a new type of minimally modified gravity theory, which also has only two gravitational degrees of freedom~\cite{DeFelice:2020eju}. 
It is called VCDM, which includes a cuscuton gravity theory 
and gives the equivalence in cosmological models~\cite{DeFelice:2022uxv}.
As shown in Ref.~\cite{Mukohyama:2019unx,DeFelice:2022uxv},
both theories are related to each other. 

As an attractive feature
in the models of the cuscuton gravity theory, it is notable that the
gravitational constant $G_{\rm cos}$ which appears in the Friedmann
equations can be different from Newton's constant $G_{\rm N}$.

In this {\it Letter}, we discuss how we can resolve the $^4$He anomaly
in the models of the cuscuton gravity theory. A modification on
$N_{\nu,{\rm eff}}$ from its standard value
$N_{\nu,{\rm eff,std}} = 3.044$ effectively has an identical effect on
a modification on the gravitational constant without changing
$N_{\nu,{\rm eff}}$ in the Friedmann equations. Thus, we can look for
a solution to the $^4$He anomaly by modifying the gravitational
constant in the cuscuton gravity theory.  In particular, we show that
we can concretely constrain the parameters in the models of the
cuscuton gravity theory from the observations.

\allowdisplaybreaks

%%%%%%%%%%%%%%%%%%%%%%%%%%%%%%%%%%%%%%%%%%%%%%%%%%%%%%%%%%%%%%%%%%%%%%
\section{Bounds from Big-bang nucleosynthesis}
%%%%%%%%%%%%%%%%%%%%%%%%%%%%%%%%%%%%%%%%%%%%%%%%%%%%%%%%%%%%%%%%%%%%%%
\label{sec:BBN}

In Ref.~\cite{Matsumoto:2022tlr}, the EMPRESS collaboration reported
the primordial mass fraction of $^4$He,
\begin{eqnarray}
  \label{eq:YpObs1}
  {\rm Y}_p = 0.2379^{+0.0031}_{-0.0030},
\end{eqnarray}
at 68$\%$C.L. According to the theoretical predictions in the BBN
computation, it gives the effective number of neutrino species,
\begin{eqnarray}
  \label{eq:NnuObs1}
  N_{\nu, {\rm eff}} = 2.41^{+0.19}_{-0.21},
\end{eqnarray}
which excludes the standard value, $N_{\nu, {\rm eff, std}} = 3.044$
predicted in the big-bang cosmology more than 2 $\sigma$.

The Friedmann equation in the $\Lambda$CDM model with vacuum energy $V_0$ is given by
\begin{eqnarray} 
  \label{eq:Friedmann1}
H^2={8\pi G_{\rm cos}\over 3}\left(\rho+V_0\right)
\,.
\end{eqnarray}
Since we can ignore the vacuum energy at the BBN stage, 
 the Hubble parameter $H$ is represented by a product of the energy density of the Universe $\rho$ and $G_{\rm cos}$ which is the
``effective'' gravitational constant appeared in the Friedmann
equations. It is notable that $G_{\rm cos}$ can be potentially different
from the Newton's constant, $G_{\rm N}$, and we may write the
difference to be $\Delta G_{\rm N} \equiv G_{\rm cos} - G_{\rm N}$.

Then, we obtain an approximate relation
\begin{eqnarray}
  \label{eq:GNnu1}
  \frac{\Delta G_{\rm N}}{G_{\rm N}} = \frac{7}{7 N_{\nu, {\rm eff, std}} + \sqrt[3]{2 \cdot 11^{4}}} \Delta N _{\nu, {\rm eff}},
\end{eqnarray}
with
$\Delta N _{\nu, {\rm eff}} \equiv N_{\nu, {\rm eff}} - N_{\nu, {\rm
    eff, std}}$,
and the prefactor of it approximately gives
$ {7}/({7  N_{\nu, {\rm eff, std} } + \sqrt[3]{2 \cdot 11^{4}}} ) \simeq
0.1343$
for $T \lesssim m_e$ with $m_e$ being electron mass. On the other
hand, we may have another value of the prefactor
($\simeq 0.1628$) in case of $T \gtrsim m_e$  with another value
of $N_{\nu, {\rm eff, std}}$ ($=3$ for $T \gtrsim m_e$). The
difference between the former and the latter values comes from that
neutrinos decoupled from the thermal bath just before $T \sim m_e$,
and only photon was heated by the $e^+e^-$ annihilation at around
$T \sim m_e$. In this study, the decoupling temperature of the weak
interaction between neutron and proton, $T_{\rm dec}$, which mainly
determines $Y_p$, tends to get delayed compared to the one in the
standard big-bang cosmology ($T_{\rm dec} \sim$~0.8~MeV) due to
$N_{\nu, {\rm eff}} < N_{\nu, {\rm eff, std}}$. Thus, we adopt the
former value (=0.1343) in this study, which also gives a more
conservative absolute value of the magnitude of
$|\Delta G_{\rm N}/G_{\rm N}|$.

From the observational data~(\ref{eq:NnuObs1}), we obtain the bound on
$G_{\rm cos}$ to be
\begin{eqnarray}
  \label{eq:GFGNBBN68}
\left. \frac{G_{\rm cos}}{G_{\rm N}} \right|_{\rm BBN} =0.915_{-0.028}^{+0.026} \quad(68 \% \text {C.L.}),
\end{eqnarray}
or equivalently,
${\Delta G_{\rm N}}/{G_{\rm N}} = -0.085_{-0.028}^{+0.026} \quad(68 \%
\text { C.L. })$.

%%%%%%%%%%%%%%%%%%%%%%%%%%%%%%%%%%%%%%%%%%%%%%%%%%%%%%%%%%%%%%%%%%%%%%
\section{Models of the cuscuton gravity theory}
%%%%%%%%%%%%%%%%%%%%%%%%%%%%%%%%%%%%%%%%%%%%%%%%%%%%%%%%%%%%%%%%%%%%%%
\label{sec:cuscutonModel}

The action of the cuscuton gravity theory is represented by~\cite{Afshordi:2006ad, Afshordi:2007yx},
\begin{eqnarray}
\label{action1}
  S &=& \int d^4 x \sqrt{-g} \Big[ \frac{1}{2} \mpl^2 R + 
  \epsilon \mu^2 \sqrt{-X}   - V(\phi)   \Big] 
  \nonumber \\
  &&
  + S_M (g_{\mu\nu}, \psi_M) \,,
\end{eqnarray}
with $M_{\rm PL}$ being the reduced Planck mass
($\simeq 2.436 \times 10^{18}$~GeV) where $R$, $V=V(\phi)$, and
$S_M (g_{\mu\nu}, \psi_M)$ mean the Ricci scalar, the potential energy
of a scalar field $\phi$, and the action of the matter field(s)
$\psi_M$, respectively. 
 $\epsilon=\pm 1$  correspond to two branches of cuscuton gravity theory~\cite{Bhattacharyya_2018}.
Here  the kinetic term  $X$ is 
defined by
\begin{eqnarray}
  \label{eq:kineticX}
  X \equiv g^{\mu\nu}\partial_{\mu} \phi \partial_{\nu} \phi\,.
\end{eqnarray} and $\mu$ is the mass parameter of the cuscuton gravity theory.
%%

%\end{widetext}

In the cuscuton gravity theory, the potential $V(\phi)$ can be
arbitrary, but if we assume the quadratic form of the potential such that
\begin{eqnarray}
V=V_0+{1\over 2}\alpha \phi^2
\,,
\end{eqnarray}
which can be regarded as the low-energy Ho\v{r}ava-Lifshitz gravity~\cite{Afshordi:2009}, we find the $\Lambda$CDM cosmology with
modification of the gravitational constant~\cite{Afshordi:2007yx}.
Here $\alpha$ means the mass squared parameter which has either a
positive or negative signature. Then, the flat Friedmann equation is
written by Eq. (\ref{eq:Friedmann1}).

Note that cosmology in the VCDM theory
 is equivalent to 
that in the cuscuton gravity~\cite{DeFelice:2022uxv}. 
We can also show the cuscuton gravity with a quadratic potential is 
 equivalent to the Einstein-aether theory~\cite{Jacobson2001} with only one coupling constant $c_2$~\cite{Bhattacharyya_2018}\footnote[1]{This model is a very special case of the Einstein-aether theory because aether field modes do not appear in the perturbation equations~\cite{Mukohyama_private}, which is consistent with the fact that the system has only two degrees of freedom.}.
In both theories (the VCDM and the Einstein-aether theory), 
we can take the Newtonian limit, which shows 
the Newtonian gravitational constant $G_{\rm N}$ is given by 
the reduced Planck mass as $M_{\rm PL}$ as  $G_{\rm N}=(8\pi M_{\rm PL}^2)^{-1}$.

In those theories, the effective gravitational constant in the Friedmann equation 
 is
given by
\begin{eqnarray}
&&{ G_{\mathrm{cos}} \over G_{\mathrm{N}}}= 
\left(1 - \frac32 \frac{\mu^{4}}{\alpha M_{\rm PL}^2}\right)^{-1}= \left(1 + \frac32 c_2\right)^{-1}=1-\frac32\beta_2
\,, \nn
&&
  \label{eq:GFGNc2}
\end{eqnarray}
where the ``potential'' of the VCDM scalar field $\varphi$, which has
the mass-dimension two, is chosen as
\begin{eqnarray}
 \label{VCDM_potential}
{\mathcal{V}}_{\rm VCDM}={V_0\over M_{\rm PL}^2}+{1\over 2}\beta_2 \varphi^2,
\end{eqnarray}
with the dimensionless mass-parameter $\beta_2$.

In order to find $G_{\rm cos}<G_{\rm N}$, each parameter should satisfy 
$\alpha<0$\,\footnote[2]{ When $\alpha<0$, the branch of $\epsilon=-1$ is required to find a  consistent 
$\Lambda$CDM expanding universe.
We would like to thank Tsutomu Kobayashi, who pointed out it.}\,, $c_2>0$ and $\beta_2>0$.
Note that 
we have the relation between these parameters as
\begin{eqnarray*}
{\mu^4\over \alpha M_{\rm PL}^2}=-c_2~~{\rm and}~~
\beta_2={c_2\over 1+\frac32 c_2}
\,.
\end{eqnarray*}

%According to the model of the cuscuton gravity theory discussed in
%Ref.~\cite{Maeda:2022ozc}, we have a relation between 
%$G_{\mathrm{F}}$
%and $G_{\mathrm{N}}$,
%with the mass $m$ of a cuscuton field appeared in its potential energy
%$V(\phi)=V_{0} - \frac{1}{2} m^{2} \phi^{2}$ with a constant term
%$ V_0$. Here, it is notable that the signature in front of
%$\frac12 m^2 \phi^2$ is negative.

In what follows, we shall first discuss
 the constraints on $c_2$ just for simplicity,
but we can translate them into the constraints on the other parameters.

From the observational bound from the EMPRESS VIII on $G_{\rm cos}/G_{\rm N}$ shown in (\ref{eq:GFGNBBN68}), we obtain the bound on $c_2$, 
\begin{eqnarray}
  \label{eq:c2BBN68}
c_{2} =   0.0620^{+0.0232}_{-0.0198} ~(68\% {\rm C.L.}), 
\end{eqnarray}
with the BBN, which gives $ 0.0235 \le c_{2} \le 0.1099$ at $95\%$
C.L. It is remarkable that $c_2 = 0$ is excluded at more than $95\%$
C.L by the BBN. Provided we assume no other change in the standard cosmology,
e.g., without assuming any change of $N_{\nu, {\rm eff}}$ (and/or
$\xi_{\nu_e}$), this may imply rejecting general relativity.

We summarize the constraints on the parameters as follows: From the EMPRESS VIII  data, we find
\begin{eqnarray*}
&&
0.0235 \le c_{2} \le 0.1099\,,
\\
&&
-42.55 \le {\alpha M_{\rm PL}^2\over \mu^4} \le 
-9.099\,,
\\
&&
\\
&&
0.0227 \le \beta_{2} \le 0.0943
\end{eqnarray*}
at $95\%$ C.L.

\begin{figure}[ht]
 \begin{center}
 \includegraphics[width=10cm]{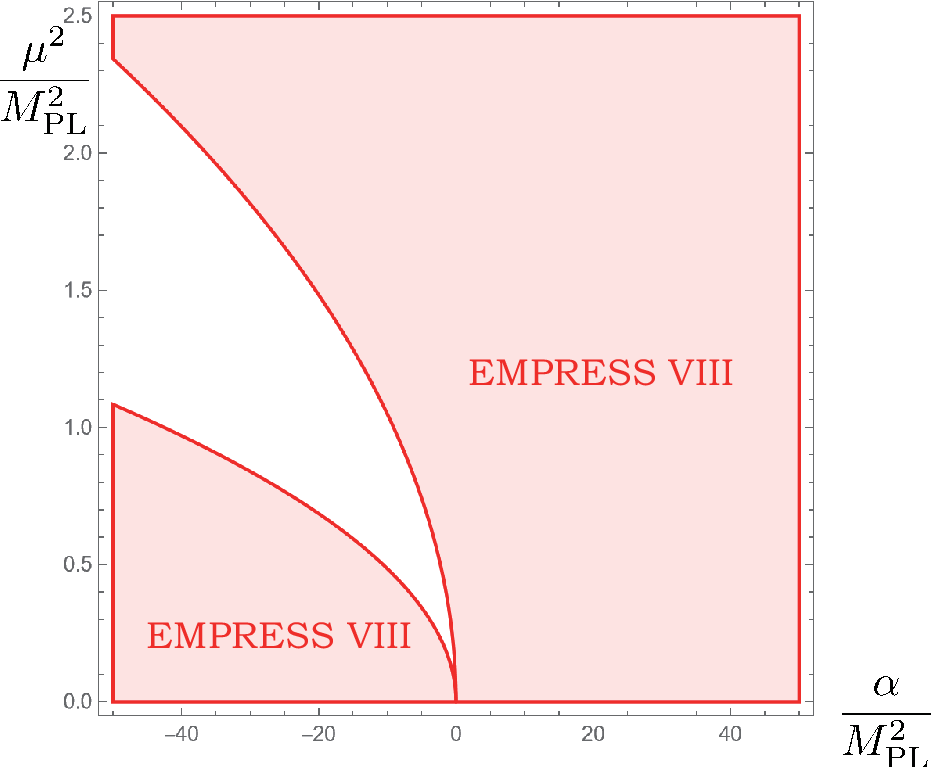} 
 \end{center}
 \caption{ Regions excluded by EMPRESS VIII with the BBN (red)
     at 95$\%$ C.L. in the $\alpha$-$\mu^2$ plane. Here $\mu^2$ and
     $\alpha$ are plotted in unit of the reduced Planck mass
     $M_{\rm PL}$.  The line of $\mu=0$, which corresponds to general
     relativity with a cosmological constant, is excluded at 95$\%$
     C.L. by EMPRESS VIII with the BBN.}
\label{fig:mualpha}
\end{figure}

In Fig.~\ref{fig:mualpha}, we show the white region allowed by EMPRESS
VIII with the BBN at 95$\%$ C.L. in the $\alpha$-$\mu^2$ plane for the
cuscuton gravity model. In other words, the red shaded regions are
excluded by observations. It is remarkable that the line of $\mu=0$,
which corresponds to general relativity with a cosmological constant,
is excluded by EMPRESS VIII with the BBN at 95$\%$ C.L.

%%%%%%%%%%%%%%%%%%%%%%%%%%%%%%%%%%%%%%%%%%%%%%%%%%%%%%%%%%%%%%%%%%%%%%
%%%%%%%%%%%%%%%%%%%%%%%%%%%%%%%%%%%%%%%%%%%%%%%%%%%%%%%%%%%%%%%%%%%%%%

One may wonder about the negative value of $\alpha$, because the
potential is unbounded from below.  However, it does not give any
instability because the scalar field $\phi$ is non-dynamical.

%%%%%%%%%%%%%%%%%%%%%%%%%%%%%%%%%%%%%%%%%%%%%%%%%%%%%%%%%%%%%%%%%%%%%%
\section{Conclusion}
\label{sec:conclusion}
%%%%%%%%%%%%%%%%%%%%%%%%%%%%%%%%%%%%%%%%%%%%%%%%%%%%%%%%%%%%%%%%%%%%%%

We have studied a cosmological model in the cuscuton gravity theory
with a quadratic potential $V=V_0+ {1\over 2} \alpha \phi^2$ to resolve the anomaly of the observational
$^4$He abundance reported by the EMPRESS collaboration.
This model is equivalent not only to the VCDM theory with a quadratic potential
with the dimensionless coefficient $\beta_2$ but also to
the Einstein-aether theory with only one coupling constant $c_2$.

About the mass squared parameter $\alpha$ in the cuscuton gravity
theory, we have obtained the allowed region as 
  $-42.55 \le ({\alpha}/{M_{\rm PL}^2})({\mu}/{M_{\rm PL}})^{-4} \le
  -9.099$,
  or equivalently $ 0.0227 \le \beta_2 \le 0.0943$ in the VCDM theory
  and $0.0235 \le c_{2} \le 0.1099$ 

in the Einstein-aether theory.
General relativity is excluded in the present approach.  Thus, this
could suggest the need for modified gravity theories such as the
cuscuton gravity theory with a quadratic potential, which can be
regarded as the low-energy Ho\v{r}ava-Lifshitz gravity.

In addition to the bound obtained by the BBN, the modification of the
gravitational constant can be also constrained by observations of
fluctuation and polarizations of the cosmic microwave background
(CMB). Here is a remark about the invalidity of a translation of a
bound on $N_{\nu, {\rm eff}}$ from the CMB observation to the one on
${G_{\rm cos}}/{G_{\rm N}}$.  Because background neutrinos are
thermally produced and have adiabatic fluctuations, a bound on
$N_{\nu, {\rm eff}}$ from the CMB is obtained by both the total energy
density and the evolution of adiabatic curvature
perturbation. Therefore, there is no simple one-to-one mapping among
the bounds on $N_{\nu, {\rm eff}}$ and ${G_{\rm cos}}$ although an
order-of-magnitude discussion would be still possible.

By using the data released by the Planck collaboration in 2018, the
authors of Ref.~\cite{Ballardini:2021evv} reported the observational
bound on ${G_{\rm cos}}/{G_{\rm N}}$ based on models of scalar-tensor
theories obtained by the CMB and the baryon acoustic oscillation
(BAO). Although we may need further detailed analysis in the present
model, their bound would be approximately applied to the current case
in the cuscuton gravity model only by an order-of-magnitude discussion
to be
$| \left. {G_{\rm cos}}/{G_{\rm N}} \right|_{\rm CMB+BAO} -1 | \sim
\mathcal{O}(0.1) \quad(95 \% \text {C.L.})$,
which would give
$- \mathcal{O}(0.1) \lesssim c_{2} \lesssim \mathcal{O}(0.1) $ at
$95\%$ C.L. This range of the error covers most of the parameter range
in (\ref{eq:c2BBN68}) allowed by EMPRESS VIII with the BBN due to its
larger error bars than those of (\ref{eq:c2BBN68}).  Thus, contrary to
the case of the BBN, general relativity ($c_2 = 0$) is even allowed
only by taking the data of CMB+BAO.

In future, $N_{\nu, {\rm eff}}$ can be measured more precisely by
planned observations of 21cm + CMB down to errors of
$\Delta N_{\nu, {\rm eff}} \sim {\mathcal{O}}(10^{-2})$~\cite{Kohri:2014hea}.
Then, we will test the gravitational constant to be a precision within
the order of
${\Delta G_{\rm N}}/{G_{\rm N}} \sim {\mathcal{O}}(10^{-3})$,
which might give a hint of quantum gravity.

We may briefly discuss a possible way to discriminate the effect of
the change in $N_{\nu, {\rm eff}}$ from the one in $G_{\rm cos}$.
When we change the $N_{\nu, {\rm eff}}$ as suggested by
~\cite{Matsumoto:2022tlr,Kawasaki:2022hvx}, there are two effects that
are measured in the CMB and 21cm observations. They are changes in
energy density and cosmological perturbation.  Therefore, precise CMB
and 21cm observations could be able to distinguish the difference
between whether the contribution is due to a change in energy density
or a change in perturbation.

There could then be a way to use the observational data to clearly
show the difference in principle if we devise a way to analyze the
data. For example, in data analysis, we propose to compare the two
cases where the CMB and 21cm data are analyzed with the fluctuation
effect in the theoretical model and the case with the fluctuation
effect cut off in the theoretical model. If there is no sizable change
in the allowed region of $N_{\nu, {\rm eff}}$ between the two, then we
can conclude that we  need the contribution of the change mainly in
energy density due to the change in $N_{\nu, {\rm eff}}$. We might
interpret this result as the change dominantly in $G_{\rm cos}$.

On the other hand, if a sizable change is produced, we understand that
the change in energy density alone cannot explain the observational
data. Thus, the change in $G_{\rm cos}$ alone does not work, which
discriminates the effect of the change in $N_{\nu, {\rm eff}}$ from
the one in $G_{\rm cos}$.

However, as far as we know such an analysis is available yet, nor have
any future experiments with sufficient sensitivity been proposed. We
hope that with the development of future observations, a suitable
experiment will be proposed that will distinguish between the two
cases.

%%%%%%%%%%%%%%%%%%%%%%%%%%%%%%%%%%%%%%%%%%%%%%%%%%%%%%%%%%%%%%%%%%%%%%
\section*{ACKNOWLEDGMENTS}
%%%%%%%%%%%%%%%%%%%%%%%%%%%%%%%%%%%%%%%%%%%%%%%%%%%%%%%%%%%%%%%%%%%%%%
We thank Niyaesh Afshordi, Tsutomu Kobayashi and Shinji Mukohyama for
useful discussions. K.M. also acknowledges 
 the Yukawa Institute for Theoretical Physics at
Kyoto University, where  the present work was completed during the
Visitors Program of FY2022.
This work was supported in part by JSPS KAKENHI
Grant Numbers JP17H01131 (K.K.), JP17H06359, JP19K03857 (K.M.), and by
MEXT KAKENHI Grant Numbers JP20H04750, JP22H05270 (K.K.). 

%%%%%%%%%%%%%%%%%%%%%%%%%%%%%%%%%%%%%%%%%%%%%%%%%%%%%%%%%%%%%%%%%%%%%%

%%%%%%%%%%%%%%%%%%%%%%%%%%%%%%%%%%%%%%%%%%%%%%%%%%%%%%%%%%%%%%%%%%%%%%

%%%%%%%%%%%%%%%%%%%%%%%%%%%%%%%%%%%%%%%%%%%%%%%%%%%%%%%%%%%%%%%%%%%%%%

%%%%%%%%%%%%%%%%%%%%%%%%%%%%%%%%%%%%%%%%%%%%%%%%%%%%%%%%%%%%%%%%%%%%%%
\end{document}